\documentclass[12pt,manuscript]{emulateapj}
\shorttitle{The cluster mass function and the universality of the IMF}
\shortauthors{Selman and Melnick}

\begin{document}


    \title{ 
	The scale-free character of the cluster mass function and the universality of the stellar IMF.
	}


\author{Fernando J. Selman   and Jorge Melnick }
\affil{European Southern Observatory, Santiago, Chile.}
   

\begin{abstract}
Our recent determination of a Salpeter slope   
	for the IMF in the field of 30 Doradus \citep{selman2005}
	appears to be in conflict with simple probabilistic counting 
	arguments advanced in the past to support observational
	claims of a steeper  IMF in the LMC field.
	In this paper we re-examine these arguments and show by explicit 
	construction that, contrary to these claims, the field IMF is expected 
	to be exactly the same as  the stellar IMF of the  clusters out
	of which the field was presumably formed.
	We show that the current data on the mass distribution of clusters themselves
	is in excellent agreement with our model, and is consistent
	with a single spectrum {\it by number of stars} of the type $n^\beta$~with $\beta$\ between -1.8 and -2.2 
	down to the smallest clusters without any preferred mass scale for
	cluster formation.
	We also use the random sampling model to estimate the statistics of the maximal
	mass star in clusters, and confirm the discrepancy with observations found by \cite{weidner2006}.
	We argue that rather than signaling the violation of the random sampling
	model these observations reflect the gravitationally unstable nature of systems with one
	very large mass star. We stress the importance of the random sampling model
	as a \emph{null hypothesis} whose violation would signal the presence of
	interesting physics.
\end{abstract}

 
 \section{Introduction}

In a recent paper \citep{selman2005} we measured
the Initial Mass Function (IMF) of the field in the 30 Doradus
super-association and found that for $7 \leq m/M_\odot \leq 40$\
the field IMF can be characterized as a power law with the \citet{salpeter1955} slope.
This result contradicts  claims of a steep
IMF for the LMC field \citep{massey1995, massey2002, gouliermis2005},
and lends support to the hypothesis of a universal IMF.
However, the observation of an initial mass spectrum of the same slope
in clusters and the field goes against
the probabilistic counting arguments of \cite{vanbeveren1982} as
interpreted by \citet[][henceforth KW2003]{kroupa2003}. Following
Vanbeveren, KW2003 posit that if the field population is entirely formed
out of disrupted clusters, then the field IMF 
must be steeper because there are many more low mass clusters
than massive ones, and low mass clusters cannot contain 
stars more massive than the clusters themselves.
  
Although an extensive review of the literature is beyond the scope
of the present work, a brief tour will place it in its proper context.
\cite{vanalbada1968b} built groups of stars by randomly sampling
an IMF $f(m)dm$\ and gives the formulas for the general order statistics
where the distribution function of the maximum mass star is one
particular case\footnote{Recently \cite{oey2005} have shown using the random sampling
model that the statistics of the maximal mass star in a number
of OB association shows evidence of an upper mass limit in the range 100-200~M$_\odot$.}.
\cite{reddish1978} gives the formulation used
by Vanbeveren, which appears to be one of the first references
that gives the formula for the mass of the maximal stellar mass
as an integral of the IMF (Equation~\ref{mmax} below).
\citet{larson1982} studied the correlation between the maximum stellar
mass and the mass of the parent molecular clouds in star-forming regions.
He noted that the observed correlation, $M_{max*} = 0.33M_{cloud}^{0.43}$,
could be explained by stochastic sampling of an IMF with Salpeter slope.
To study dynamical biasing \citep{vanalbada1968a} in binary star formation \cite{mcdonald1993} used a two-step
process in which they sample stars assuming that a certain fraction $g(N)$\ of them come
from groups of size N, and then sampled a stellar mass spectrum to build several statistics
of binary stars. The method  was extended by \cite{sterzik1998} to study the decay of
gravitational few body systems. To build
their clusters they introduced what they called a ``two-step" approach which later lead
them to the ``two-step initial mass function" \citep{durisen2001}: first draw a cluster
mass from a cluster mass-function, then draw enough stars from  an stellar IMF to
add to the cluster mass. The method presented here is similar, but with the
important difference that we do not censor by mass, but we rather work with a cluster
spectrum by number, and draw stars from a stellar mass function. This has the important
consequence that we know by construction that there are no preferred mass scales other
than those present in the stellar mass spectrum or the spectrum of clusters by number.
A similar model was used by \cite{oey2004}\ to study the distribution of clusters
by numbers in the SMC to conclude that the data for the high mass groupings studied
is consistent with an $n^{-2}$\ distribution.

\cite{vanbeveren1982} using the then assumed Salpeter slope
for the field stellar IMF concludes that \emph{``massive aggregates would
contain more OB type stars than predicted by the Salpeter IMF."}
Interestingly, KW2003 turn the argument around and use the well established Salpeter
form for the cluster stellar IMF for $m>1 M_\odot$~to infer a slope
steeper than Salpeter for the field stellar IMF in the same mass range.
With the exception of the LMC work mentioned above, the evidence
for a Salpeter slope for the field stellar population is overwhelming,
from the original \citet{salpeter1955} work on the Milky Way
to more recent work such as that of \cite{scalo1986}
that steepen slightly the slope of the high mass end from 2.35 to 2.7
\citep[for recent reviews on this topic the reader is referred to][]{kroupa2002,elmegreen2006}.
Recently, Weidner and Kroupa (2006; henceforth WK2006) used an extensive
set of Monte Carlo simulations to investigate the question of whether clusters
could be constructed by sampling stellar IMFs using different sampling prescriptions.
Their strong conclusion is that the model in which clusters are 
built by random sampling of a (Salpeter) stellar IMF is falsified by the statistics
of the maximal mass star in clusters, stating: \emph{``With this contribution
we demonstrate conclusively that the purely statistical notion is false,
and that the stellar IMF is sampled to a maximum stellar
mass that correlates with the cluster mass."} 

The purpose of this paper is to revisit this issue and to examine the question of whether clusters
and the field sample a universal stellar mass distribution.
We show that what really matters
is the (poorly determined) cluster mass spectrum in the range of single star masses ($M<150M_{\odot}$), and that
it is more natural to work with the cluster ``number of stars spectrum", $P(n)$, 
the probability of a cluster having $n$ stars.
We address this question in two different ways: by studying it from first principles, and by actually
doing Monte Carlo experiments building clusters  randomly
sampling a universal stellar IMF and comparing the results with the observations. 
The random sampling model has no other physics in it than that input from the stellar mass
spectrum and the cluster number spectrum. It should be considered as a \emph{null hypothesis}
for interesting physical processes: its violation signals the presence of interesting physics.
Occam's razor should be used with all models that violate the \emph{null hypothesis} until
strong observational evidence renders the model untenable.

In Section~\ref{formalism} we present the formal framework for the subsequent analysis,
and give an analytical parametrization of the stellar IMF
that agrees reasonably well with observations at all masses. In that section we also
present an analytical relationship between the cluster mass function and the stellar IMF. We use this
relation to conduct Monte Carlo experiments to simulate
the mass distribution of  clusters.   In Section~\ref{obs} we compare our simulations with the observed 
distribution of embedded clusters presented by \citet[][~henceforth LL2003]{lada2003}. The claim by LL2003
that there is a preferred mass scale for cluster formation is not born out by
our analysis, and a critical discussion to uncover the sources of this
discrepancy is presented. In Sections~\ref{compare} and~\ref{disc} 
we challenge the view that all stars 
form in clusters and argue that our results favor a view where stars form, or at least acquire their
final properties, before cluster formation.
Section~\ref{summary} summarizes our results and ends with the usual plea for more observations.

 \section{Building a field population from clusters: the formalism.}
\label{formalism}

We will use the term population  in
the statistical sense: a
set with infinitely many elements \citep{brandt1998}.
Consider a population of stars with a Salpeter frequency
distribution of masses $f(m)$.  The mass $m$\ 
of the stars therefore is a random variable with a frequency
distribution $f(m)$. Let us draw samples from such population
with a fixed number of stars $n$\ and frequency distribution $P(n)$\footnote{
In this paper the symbol $P(x)$\ stands for probability when
the random variable $x$ is a number (i.e. $n$), and for a  the frequency
distribution when $x$ is a mass (i.e. $m; M; m|M$). The meaning should be clear from the context.
}.
Each of the samples will be called a \emph{cluster} although such ``clusters'' can contain a single star. 
This construction is analogous to those used in previous
work studying the properties of HII regions in galaxies
\citep{oey1998}, the more general study of Poissonian
fluctuations in population synthesis models by \citet{cervino2002},
and the analysis of the isolated massive stars in the Milky Way by
\cite{dewit2005}.

The frequency distribution function of cluster masses will be given by,
\begin{eqnarray}
\xi_{cl}(M) = \sum_{\small n=1 \atop M = m_1 + m_2 \cdots + m_n}^\infty \mathcal{F}_n(m_1,m_2,\cdots,m_n)P(n) ,
\end{eqnarray} where $\mathcal{F}_n$\ is the multivariate frequency distribution of
masses for a sample of size $n$, and the summation is understood
also as a multiple integral over all masses $m_1,\cdots,m_n$\ satisfying the
constraint that they add up to $M$ (which imply quite a complex domain of integration).
If the sample is  random then the following two
conditions are satisfied: 

\begin{itemize}
\item[(a)] the individual $m_i$\ must be independent, that is,
\begin{eqnarray}
\mathcal{F}_n(m_1,m_2,\cdots,m_n) = f_1(m_1)f_2(m_2)\cdots f_n(m_n),
\end{eqnarray}

\item[(b)] the individual marginal distributions
must be identical and equal to the frequency distribution
of the parent population, that is,
\begin{eqnarray}
f_1(m_1) = f_2(m_2) = \cdots = f_n(m_n) = f(m).
\end{eqnarray}
\end{itemize} We can write an explicit expression for the cluster
mass function (we will use lowercase for stellar quantities and
uppercase for cluster quantities). Because we consider only random
samples, the variable $M = m_1 + m_2 + \cdots + m_n$\ is also a random
variable.  Thus, the distribution function of $M$, $F_n(M)$, can be written as
\begin{eqnarray}
F_n(M) = \int\limits_{\small -\infty\atop {\large M < m_1 + m_2 + \cdots + m_n < M + dM} }^{+\infty} f(m_1)f(m_2)\cdots f(m_n)dm_1\cdots dm_n.
\end{eqnarray}
We have used a somewhat unusual notation under the integral sign to indicate that  the domain of integration is restricted to total masses
between $M$ and $M+dM$ only. We can write this condition on the total masses as a Dirac delta-function in terms of its Fourier expression \citep{morse1953}
\begin{eqnarray}
\delta(M-\sum m_j) = {1\over 2\pi} \int_{-\infty}^{+\infty} e^{-i(M-\sum m_j)t}dt.
\end{eqnarray}This allow us to integrate over all positive $m_j$\ and thus to avoid the problem
posed by the difficult domain of integration:
\begin{eqnarray}
F_n(M) & = & {1\over 2\pi} \int_{-\infty}^{+\infty} e^{-i(M-\sum m_j)t}dt f(m_1)f(m_2)\cdots f(m_n)dm_1\cdots dm_n \\
	& =  & {1\over 2\pi} \int_{-\infty}^{+\infty} e^{-iMt} \prod_{j=1}^n e^{-im_jt}f(m_j)dm_jdt\\
	&  = & {1\over 2\pi} \int_{-\infty}^{+\infty} e^{-iMt} \phi^n(t)dt
\end{eqnarray} where $\phi(t)$\ is the characteristic function of $f(m)$, that is,
the Fourier transform of the probability density:
\begin{eqnarray}
\phi(t) & = & \int_{-\infty}^{+\infty} e^{imt}f(m)dm.
\end{eqnarray}Thus,
\begin{eqnarray}
F_n(M) & = & {1\over 2\pi} \int_{-\infty}^{+\infty} e^{-iMt}\phi^n(t)dt.
\end{eqnarray} Finally, the cluster mass function (Equation 1) can be written as,
\begin{eqnarray}
\label{eqMC}
\xi_{cl}(M) & = & \sum_{n=1}^\infty F_n(M)P(n),\\
\xi_{cl}(M) & = & {1\over 2\pi} \int_{-\infty}^{+\infty} e^{-iMt}\underbrace{\sum_{n=1}^\infty P(n)\phi^n(t)}_{\Phi(t)}dt
\end{eqnarray}were $P(n)$\ is an arbitrary probability distribution of the number of stars in clusters. We see above that
the characteristic function of the cluster mass function, $\Phi(t)$, and the characteristic function of the stellar
mass function, $\phi(t)$, are related as,
\begin{eqnarray}
\label{charfunc}
\Phi(t) = \sum_{n=1}^\infty P(n)\phi^n(t).
\end{eqnarray} 

Equations~\ref{eqMC}-\ref{charfunc} form the basis for either an
analytical, or a Monte Carlo approach to the statistical
simulation of clusters. Its importance resides in that it relates
the cluster ``number of stars"  distribution function, $P(n)$, the (universal)
stellar IMF, and the actual cluster mass function. We will study a simple
analytical case to illustrate its properties 
and proceed with full Monte Carlo simulations.
Consider for example the case in which the cluster stellar mass function is simply
$f(m) = \delta(m-m_*)$, that is, a cluster
with a single stellar mass species of mass $m_*$. In this case $\phi(t) = e^{im_*t}$,
and $\phi^n(t) = e^{inm_*t}$\ from which we
obtain $F_n(M) = \delta(M-nm_*)$\ as expected.
More generally, we can use the relationship between
cumulants and moments of a distribution \citep[][p.69]{kendall1977} to determine that
the mean mass of $F_n(M)$\ scales with $n$, and its width
scales with $\sqrt{n}$.

We should notice that  we have constructed a set of clusters
with strictly the same mass spectrum as that of a field
built by their total destruction.
Since we can set $P(n)$\ to be any function, and in particular $P(n)=0$\ for $n<N$\ for an arbitrary $N$,
this important result  holds independently of the lower cut-off in the cluster
number spectrum:  \emph{the stellar mass spectrum of clusters and of a field built entirely out 
of disrupted clusters can be strictly the same.} At first sight this may seem to be an almost trivial result, but
notice the subtlety revealed by the following {\it gedanken} experiment:  sample a universal stellar IMF to create a sample
of $n$\  clusters with different numbers of stars according to $P(n)$ and partition this sample 
according to their mass $M$ to determine the stellar IMF conditioned to the parent cluster mass, $P(m|M)$
(the probability that a star has mass $m$ if the parent cluster mass is $M$).  For $M$ in the range of
single star masses, this probability is 
not independent of $M$ and one would get the impression
that the stellar IMF does depend on the cluster mass, that is, it is not universal. However, we know  by construction that the
clusters have their stars drawn from exactly the same stellar IMF; what depends on
cluster mass is the {\it conditional probability}. The purpose of this paper is to
investigate whether the observations are in agreement with the $P(m|M)$\ that derives from the random
sampling model, or whether they falsify it. Furthermore, we know from probability theory
\begin{eqnarray}
\label{conditionalP}
P(m) = \int_0^{+\infty}P(m|M)P(M)dM
\end{eqnarray}that in the above construction we should recover the input stellar mass function. Thus, for given $P(n)$ and $P(m)$,
$P(m|M)$\ and $P(M)$\ must  {\it conspire} so that Equation 14 is satisfied.
Using this relation one finds for the simple case of the single mass species
clusters with $n$ stars that $P(m|M)=\delta(M-nm_*-m+m_*)=\delta(m-m_*)$, independent of 
$M=nm_*$, as it should be. 

As it is shown in Section~\ref{compare}, this ``conspiracy" is not present
in other treatments of this problem, where $P(m|M)$\ and $P(M)$\ are taken
to be totally independent. Because of this it makes more sense to
work with the cluster ``number function", $P(n)$.

Equations~\ref{eqMC}-\ref{conditionalP} are the fundamental relations relating
the stellar and cluster mass spectra in the
random sampling model. Given the stellar mass function $f(m)$\ and $P(n)$\ they fix the form
of the cluster mass function $\xi_{cl}(M)$\ and of $P(m|M)$, which can then be compared with observations.
 
\subsection{Monte Carlo simulations}
\label{mc}

Using the formalism described in the previous section we build clusters by randomly sampling the following ``universal'' stellar IMF,
\begin{eqnarray}
dN\propto {m^\alpha e^{-(m/m_2)^q}\over (m_1^2+m^2)^{\gamma/2}}dm
\end{eqnarray}
\noindent where $\gamma$, $\alpha$, $m_1$, $m_2$~and q~are chosen to give the appropriate
behavior at low and high masses: $-1+\Gamma = \alpha-\gamma=-2.35$, $m_1=0.3\/M_\odot$,
$m_2=150\/M_\odot$, $q=3$. Figure~1 shows this analytical
stellar IMF together with the IMF of the Trapezium cluster \citep{hillenbrand2000,luhman2000,muench2002}.
Our analytical formula departs from the observations at the higher masses  because we have chosen
to preserve the Salpeter slope for $1 \leq m/M_\odot \leq\ \sim\!\!120$.
Henceforth we will call this function the Salpeter IMF.

\begin{figure}
   \includegraphics[width=6cm,angle=-90]{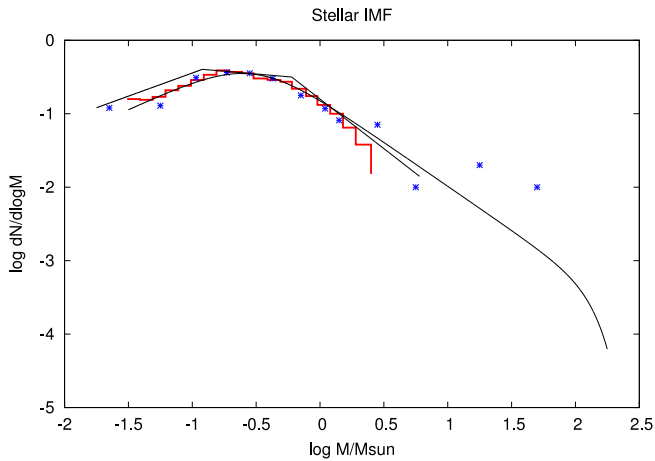}
      \caption{
        The analytical stellar IMF that we use for our MC
        experiments compared with the IMF of the Trapezium
        cluster by \citet{hillenbrand2000}, steps; \citet{luhman2000},
	segmented solid lines;
	and \citet{muench2002}, asterisks.\label{stellarIMF}}
\end{figure}

We sample the Salpeter IMF to create clusters with \emph{n} stars assuming 
a scale-free frequency distribution $P(n)\propto n^\beta$,
and we build the cluster mass spectrum using Equation 10,
\[
\xi_{cl}(M) = \sum_{n=1}^{\infty} F_n(M)P(n),
\] where $F_n(M)$~is the mass distribution function of clusters
with exactly $n$\ stars. Notice that this process is scale-free only if the sum starts from $n=1$,
in which case the only mass scales of the problem are $m_l$\ and $m_u$, the lower and upper
mass cut-offs of the stellar IMF.
Our Monte Carlo simulations consist of
repeatedly drawing \emph{n} stars from the Salpeter IMF,
calculating $M=m_1+\cdots+m_n$\ as the mass of a cluster
with \emph{n} stars, and then obtaining $F_n(M)$.

For the value of $\beta$\ there have been a multitude of studies of massive clusters
in galaxies, which gives for the mass functions values ranging between $\beta = -1.85$
\citep{degrijs2006} to $\beta = -2.4$\ \citep{hunter2003}. More extensive references
are given in \cite{elmegreen2006}. For smaller clusters \cite{dewit2004,dewit2005} claim
that their data on isolated massive star formation can be understood if $\beta=-1.7$. 
For massive clusters one can directly use the same exponent for the mass function as
for $P(n)$ because, as discussed above, the total mass scales with $n$ and the width 
for fixed $n$ scales with $\sqrt{n}$. In this work we will explore $\beta = -1.8, -2.0$, and $-2.2$.

\section{Comparison with observations}
\label{obs}  

We have identified two observational tests that can be performed
to check the validity of our \emph{null hypothesis.}
First, we will see if we can reproduce
the form of the embedded cluster mass function; second, we will see
if we can reproduce the statistics of the most massive star
in clusters. We are aware that we are leaving out tests regarding
the characteristics of small $n$\  multiple systems, which could
falsify it\footnote{The study of the statistics of small n multiple
systems is beyond the scope of the present work, but even here where
observations of the frequency of high mass doubles appear to violate
the simple random sampling model, there are physical mechanism which
explain them preserving the model, namely,
dynamical biasing\citep[see][and references therein]{sterzik1998}.}.
But multiple systems, although numerous, are not the
main source of stars in the field, so they will not affect the
main conclusions of the present work, namely that the stellar
and cluster field IMF can be the same.
 
\subsection{The embedded cluster mass function}

Due to the difficulty defining unbiased complete samples,
the important range of clusters masses in the regime of
stellar masses is not well studied. There are nevertheless
two relatively recent sources based on extensive surveys of the
literature at the time of publication: \citet{porras2003} and \citet{lada2003}.
We prefer to use LL2003 four our analysis because they give estimates of the masses of the clusters, although 
only 4 of the clusters in the Porras et al. list that satisfy the
constraint on minimum number of stars of LL2003 are not included in this catalog. The cluster masses given in LL2003 were obtained by
modeling source counts as a function of limiting
magnitudes for two model clusters with ages of 0.8~Myr and 2~Myr,
corresponding to the ages of the Trapezium and IC~348 clusters
respectively. They assumed a universal IMF and used the average
of the mass determined for the two assumed ages. 

Figure~2 shows the empirical data of LL2003
together with the results of six runs of our MC experiments
in which we built clusters with the above $P(n)$\ for $n\geq35$, drawing
72 clusters at the time (the parameters of the observations of LL2003).
We note the excellent agreement between the simulations
and LL2003 except for the mass bins at $\log M\sim0.95$
and $\log M\sim3.5$\ which are totally de-populated in LL2003.
For $\beta=-1.8$\ the smallest mass bin is de-populated in 15\% of our
simulations while in 70\% of the simulations contains 2 clusters or less.
The highest mass bin is populated in only $\sim40\%$
of the simulations and in almost 100\% of the simulations
contains less than 2 clusters.
LL2003 proposed that the downturn at smaller masses was
evidence for a favored cluster formation mass scale at around $M\sim 50M_\odot$.
However, our simulations indicate that this downturn is naturally explained
by the cutoff in \emph{n} they introduced in an otherwise scale-free spectrum, 
without the need
to invoke a special cluster formation scale. The figure shows that the data is
best modeled if the cutoff in $n$\ is a bit larger than the LL2003
criterion of $n>35$\ to select clusters. This is probably the effect of
having a sample with an inhomogeneous magnitude limit so that $n>35$\ 
becomes only a lower limit to the actual cutoff.

\begin{figure}
   \epsscale{1.10}
   \plotone{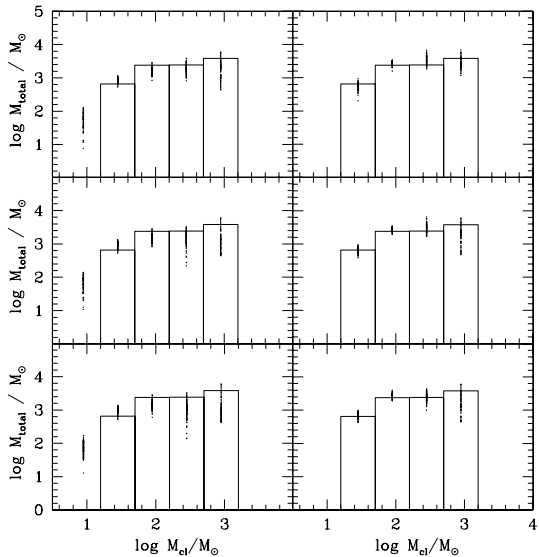}
    \caption{\label{clusterIMF}
        The Lada and Lada (2003) spectrum of masses of embedded
        clusters together with the results of Monte Carlo
        simulations in which we sample the stellar IMF
        with a number probability distribution $\beta={-1.8}$,
	top row; $\beta={-2.0}$, middle row; $\beta={-2.2}$, bottom row.
        For $n\geq35$, left column; and $n\geq70$, right column. 
           }
\end{figure}

\subsection{The statistics of the maximal mass star in clusters}
\label{compare}

\begin{figure}
   \epsscale{1.2}
    \plotone{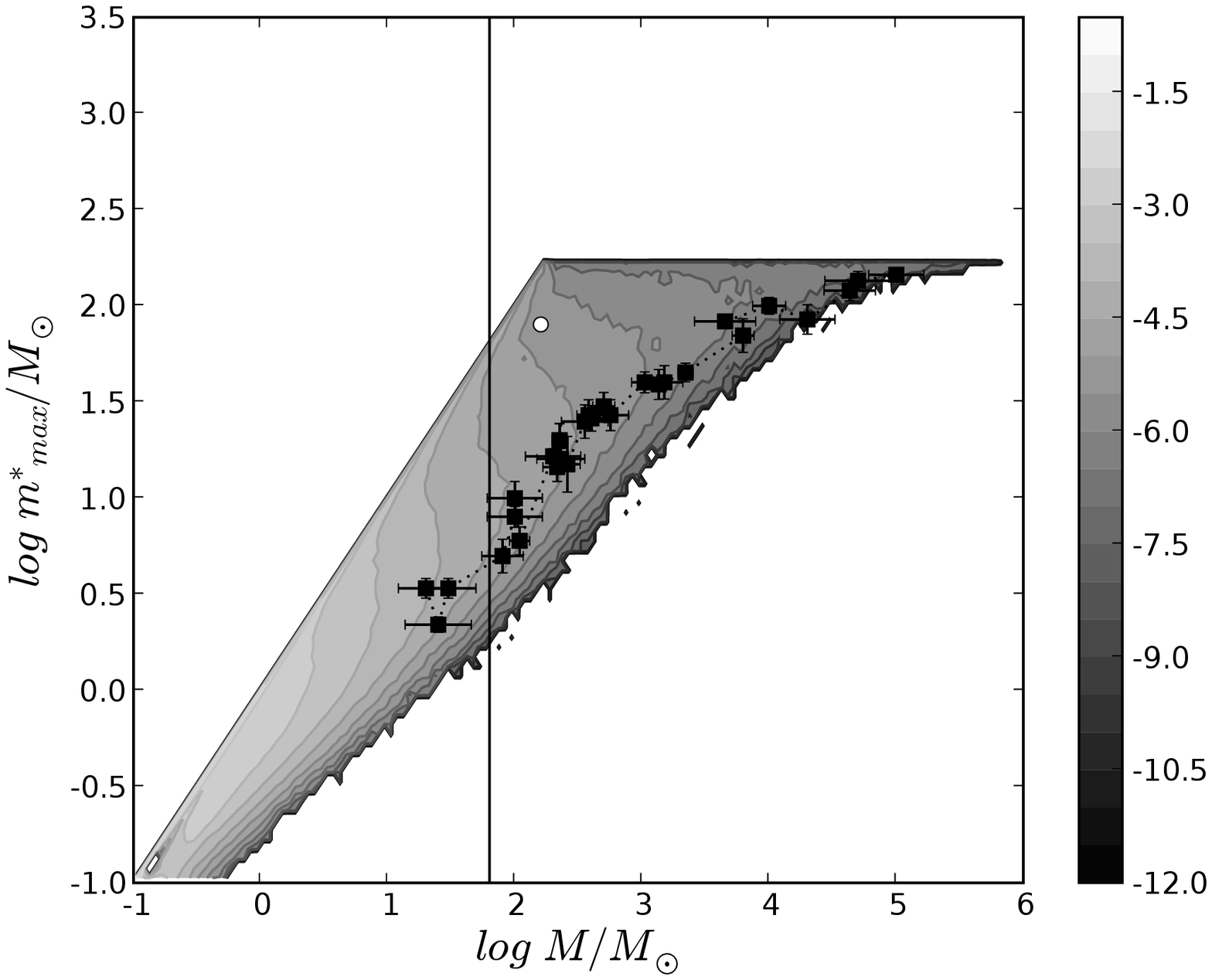}
    \plotone{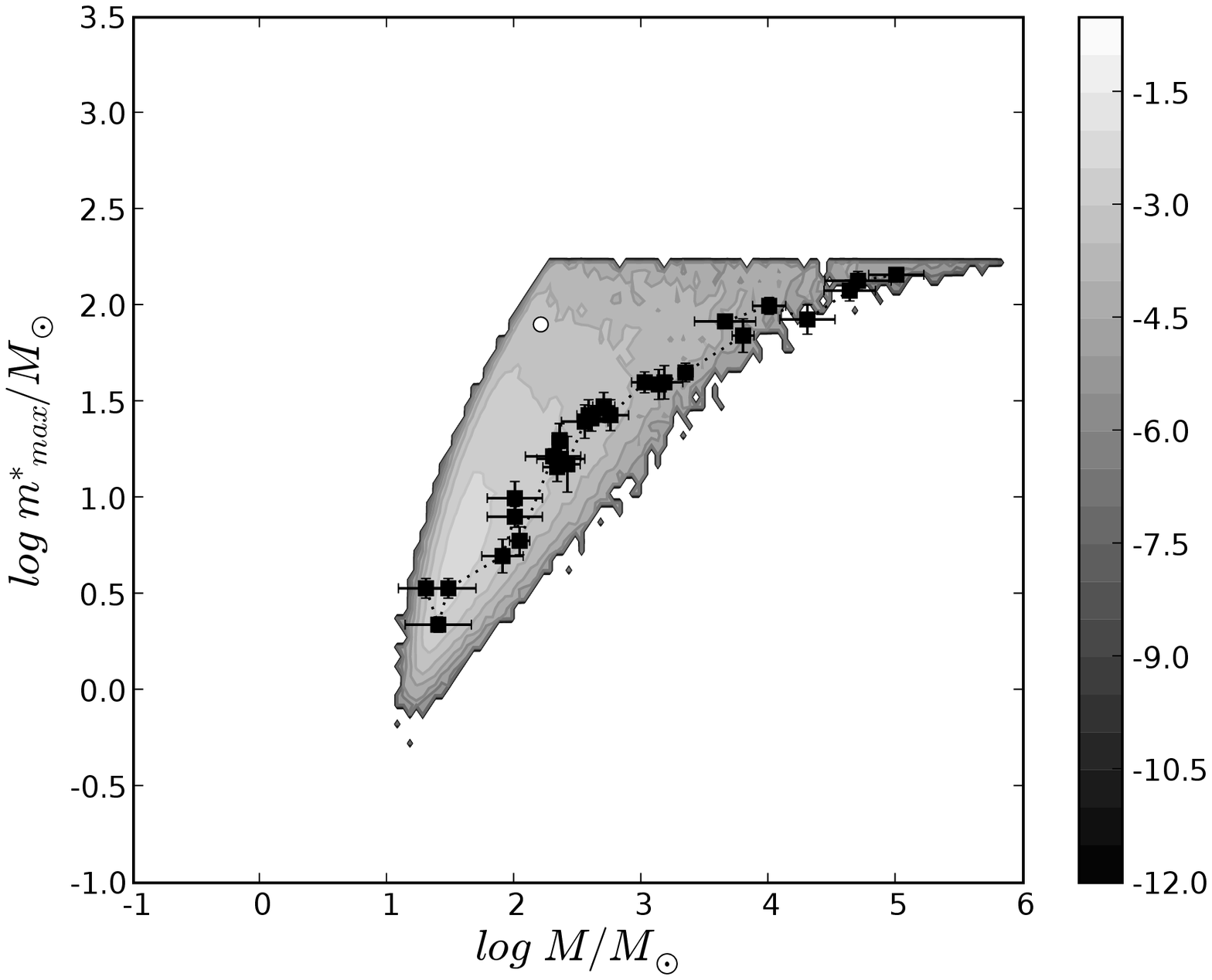}
    \caption{\label{bivariate}
	(a, top) The bivariate probability distribution function of
	maximum stellar mass and cluster mass.  The overlaid points correspond to the
        data in \cite{weidner2006}. This figure show the result with a cluster $P(n)\sim 1/n^2$
	starting at n=1. The grey levels correspond to the bivariate probability of finding
	a cluster in a log-mass bin of size $0.5\times0.1$. The white circle represents WR20a
	taken as a system. The points represent the data points in WK2006 with a few additions
	from Weidner (2007). The vertical line is drawn at a log M value of 1.8.
	(b, bottom) Same as (a) but with $n>50$. For details see main text.
}
\end{figure}

WK2006 argued that their observed correlation between the maximal star mass
and the total mass of clusters is not consistent with the hypothesis that
clusters are formed by random sampling of a universal stellar IMF.
WK2006's sample of clusters is strongly affected by a
size-of-sample effect (see Appendix~\ref{ap:sizeofsample}).
Because of the impracticality of finding a large ensemble of
small clusters and thus avoid the problems introduced by the
size-of-sample effect, WK2006 performed Monte Carlo experiments to determine
the statistical properties of each of their  three
sampling methods. We were puzzled by their Figure~3, which shows that
for their random sampling method (that corresponds to our Monte Carlo models),
the curves of maximal mass star versus cluster mass ($M_{ecl}$) have two maxima
in the range between  $\sim 25M_{\odot}$\ and $\sim 250M_{\odot}$. For example, for $M_{ecl}=100M_\odot$\ the curve peaks
at $\sim10~M_\odot$~and then again at $\sim100~M_\odot$.  Since it is precisely in this mass range that the model curve departs most
strongly from the data points, we thought that the double peaks could be the result of a computational error.
We therefore decided to repeat the calculations using our independent algorithm
to build the bivariate (maximal stellar mass -- cluster mass) probability
distribution. The results of our simulations are shown Figure~\ref{bivariate}, where,
much to our surprise, we reproduce the double peaks obtained by WK2006!

Figure~3 shows the results of many MC experiments of the random sdampling model
from which we have calculated the bivariate probability distribution to have a
cluster in a given log-mass bin of width 0.5, with a star of maximal mass in a
log-mass bin of width 0.1. Lighter areas correspond to a higher probability.
Figure~3a  correspond to MC experiments in which anything is considered a cluster, even
system with n=1. Figure~3b considers only cluters with n$>$50. The vertical line
in Figure~3a marks the position of $\log M = 1.8$. Notice that as one moves from
bottom to top along this line one will cross contour levels that at first increase
until the bivariate distribution reaches a maximum at $\log m^*_{max}\approx 0.9$.
If one continues to move it will reach a minimum at approximately
$\log m^*_{max}\approx 1.4$, and then it starts increasing again reaching a maximum
at the point in which all the cluster mass is in a single star.
Notice that this double (local) maxima feature comes from the nature
of the probability distribution of star masses conditioned to cluster
mass ((Fig~\ref{fig:conditionalP}, see below).

Interestingly, the clusters from the compilation
of \cite{weidner2006} (crosses in Fig.~\ref{bivariate}) all concentrate in the ridge of the
distribution defined by the first (lower mass) peak described above and do not cover
the full mass range allowed by our models. To a lesser extent this is also true of the models by
\cite{weidner2006}, whose Figures~4 and 5 show the data to have a much smaller
dispersion around the mean than the models. Nevertheless, the random sampling method
deviate most from the data due to its ``double peaked'' mass distribution.
With our preferred random sampling method clusters with masses in the stellar
mass range the most massive stars can have masses similar to the total cluster mass.
Moreover, the conditional probability distribution of stellar masses, $P(m|M)$,
for clusters in the stellar mass range shows that some clusters can be dominated
by one or at most a few high mass stars (Fig~\ref{fig:conditionalP}).
This increase in the probability distribution of stellar masses near the cluster mass
is also visible in Figure~2 of \cite{durisen2001}, so the critical
questions are whether this effect is real, and if so, whether it is significant.
The fact that the effect is seen in three independent investigations argues strongly
for the reality of the peak in the random sampling model, but
its significance is debatable. On the one hand, the effect arises
from partitioning the data into mass bins, and it is forced into existence
by the need to satisfy Equation~\ref{conditionalP}, so it is of no physical significance.
On the other hand, it biases (toward large values)
the mean maximal stellar mass versus cluster mass curve used
by \cite{weidner2006} to falsify the random sampling model, so it is
highly significant.

Is this a real violation of our \emph{null hypothesis} signaling the presence of
some interesting physical effect or is it the effect of improper data, or its analysis,
or both?  Although WK2006's conclusions are based on a very limited data-set\footnote{For example, 
their favored sorted sampling model predicts no single star clusters at all, while 
\citet{dewit2004, dewit2005} find truly isolated massive stars of spectral
types ranging between O5 and O9 \citep[see also][]{zinnecker2007}. None of these ``single star clusters'' are included in KW2006.},
it is unlikely that this alone can explain the difference in the distribution of
the data points and that predicted by the models: the number of clusters in each mass
bin is small but the total sample is not that small and at all masses the data
points are delineating the lowest maximal stellar mass \emph{ridge} of the distribution.

\begin{figure}[t]
   \epsscale{1.2}
    \plotone{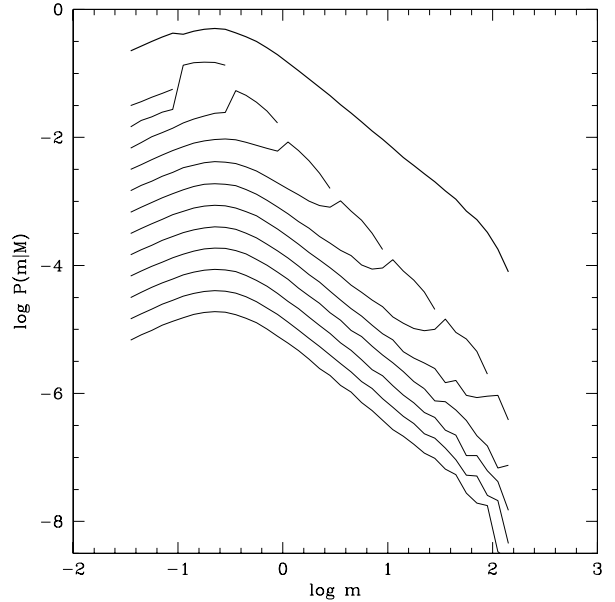}
    \caption{\label{fig:conditionalP}
        Conditional probability distributions of stellar masses conditioned
        to cluster mass. The topmost curve correspond to P(m) where the
	cluster mass has bin marginalized. The other curves correspond from top to bottome
	to P(m|M) for log M equal to -1.25, -0.75, -0.25, 0.25, 0.75, 1.25, 1.75, 2.25,
	2.75, 3.25, 3.75, and 4.25 respectively. 
    }
\end{figure}

One possible explanation could come from the highly hierarchical nature of
young stellar systems and the somewhat arbitrary way in which the parent
object of the maximal star is chosen. For example, the cluster Westerlund~2 has the well
known massive binary WR20a, both components of which with masses $\sim 80 M_\odot$\ \citep{rauw2004, rauw2005}.
Considering that the ratio of its components separation to the cluster size is smaller
than the ratio of the cluster size to that of the Milky Way, there is no objective reason
not to have such binary star as a single point in Figure~\ref{bivariate}, in which
case we would have a data point in the area devoid of points in that Figure. An objective
algorithm to identify clusters of different number of stars is needed. A step in that
direction is the work of \cite{oey2004} which used the algorithm by \cite{battinelli1991} to
identify groupings of stars and showed that the distribution of clusters by number of stars
is consistent with $n^{-2}$\ down to single stars. What is needed is then to determine
the maximal mass star and total mass of those clusters to build the bivariate probability
distribution.

Finally, another explanation is that these data do indeed falsify our \emph{null hypothesis} and
that they signal the presence of some interesting physical effect that results
in the mass of the most massive star to depend on the mass of its parent cluster.
One possibility could be that stars form in an ordered fashion, less massive stars
forming first. Once a high mass star is formed this one cleans the cluster
of its placental material and the cluster rapidly disintegrate \citep{elmegreen1983}. This is the explanation
favoured by WK2006.  But there is another possibility: clusters with maximal mass star of too
large masses can be more gravitationally unstable. A cluster with a maximal mass star of very
large mass is characterized also by having a smaller total number of stars, and its conditional stellar mass function
is flatter (see Figure~\ref{fig:conditionalP}). \cite{terlevich1987} found that a flatter
mass spectrum results in a considerably smaller half-life: her model XV evolves an order
of magnitude faster that a cluster with a normal Salpeter IMF. This explanation has
the feature that it does not violate the \emph{null hypothesis,} because clusters are
formed according to the random sampling model but their lifetimes, and thus the probability
of observing them, depend on the mass of the maximal stellar mass in them.

\section{Discussion}
\label{disc}

As mentioned in the Introduction, our results depart radically from those of KW2003.
Their conclusion that the field built by clusters must have a steeper stellar IMF
can be criticized on several accounts as follows.
To begin, as noted by \citet{elmegreen2006},  for the observed range
of cluster masses, the predicted steepening in the stellar IMF is rather small going
from $\Gamma = -1.35$\ in clusters
to $\Gamma \sim -1.5$~in the field (Figure~2 in KW2003).
This is well within the observed variance in the
stellar IMF of clusters, and can be comfortably ascribed
to systematic errors in the determination of the slope \citep[see e.g.][]{kroupa2002}.

A second problem is that KW2003 implicitly assume that the cluster mass spectrum
is well determined down to masses comparable to those of the
smaller stars and is approximately given by
$\xi_{cl}\sim M^{-2}$\ for $5 \leq M/M_\odot \leq 10^7$.
As mentioned in the previous section, however, LL2003 claim that the cluster mass spectrum turns
abruptly down for masses below $M \sim50M_\odot$, that is, it appears to have a
preferred mass-scale. This reduces the number of small mass
clusters thus reducing the predicted difference  between
cluster stellar IMF and the field.
Almost paradoxically, our model invoking a universal stellar IMF
shows that this preferred mass-scale is most likely not a physically significant
feature of cluster formation!

Another problem is that KW2003 use a Procrustean approach in their modeling of clusters
where all clusters are forced to have one \emph{maximal mass star}, that
is, a star of the maximum mass allowed by the stellar IMF.  This assumption,
\begin{eqnarray}
\label{mmax}
1 = \int_{m_{max*}}^{m_u} f(m)dm,
\end{eqnarray}
forces the upper mass cut-off of the IMF to be an increasing function of cluster mass,
varying as $M^{1/x}$. From the discussion leading to Equation~\ref{conditionalP} it is clear
that we can not recover the input IMF unless we include in our sums clusters for which $m=M$.

\citet{elmegreen2006} expanded the formalism of Vanbeveren and showed analytically and
numerically that for a power-law mass distribution of clusters of slope $\beta\leq 2$,
the summed IMF  of a population of clusters  is indistinguishable from  the cluster IMF.
However, the functional form of the conditional probability $P(m|M)$ adopted by Elmegreen
{\it does not} satisfy Equation~\ref{conditionalP} for any value of $\beta$; it just happens
that for $\beta\leq 2$ the summed IMF is {\it almost} the same as the cluster IMF (they differ
by a logarithmic multiplicative factor). This clearly shows that the finding of KW2003, 
that for $\beta>2$ the summed IMF becomes significantly steeper than the individual cluster IMF, 
arises from the assumption that, besides a normalization factor that depends on the cluster mass 
through Equation~\ref{mmax}, $P(m)$ and $P(m|M)$\ have the same functional form, (simple power-laws 
of the same slope in the case of Elmegreen).  While this is a very good assumption for very massive 
clusters, it clearly does not apply for clusters in the stellar mass range which are the ones responsible 
for ``tilting'' the sum-IMF for $\beta>2$. Thus, even within the Vanbeveren formalism the IMF of clusters 
and the field can be strictly the same for any cluster mass distribution.

The results of WK2006 are confirmed in this work in the sense that we also find that
real clusters cover a significantly smaller part of the maximal stellar mass and cluster mass
space than  permitted by the random sampling model. WK2006 go then to modify their sampling
algorithm in such a way as to reduce the probability of having very large mass stars in their
clusters arguing that \emph{``Star clusters appears to form in an ordered fashion, starting
with the lowest-mass stars until feedback is able to outweigh the gravitationally
induced formation process.''} Although this is a possible scenario we favour a different
one in which the area of the bivariate distribution allowed by the random sampling
model is rendered unstable due to the extreme nature of the mass spectrum therein,
in accordance with the simulations of \cite{terlevich1987}.
How much ``trimming'' of the bivariate distribution can be actually accomplished this
way will be the subject of a future work.

Our strong conclusion is that the observations of the stellar IMF and the mass spectrum of young 
clusters are consistent with the hypothesis that clusters form by random sampling of a universal 
stellar IMF.  This conclusion leads us to challenge the received hypothesis that clusters are the fundamental 
building blocks of the stellar populations in galaxies \citep{dewit2004,dewit2005}.  In this view clusters are 
given an independent existence from before the time that stars form. In our view, following the ideas of 
Elmegreen (1997), stars form in giant molecular
clouds in a hierarchy of structures with different numbers 
and masses.  Some of these structures end up forming large clusters 
(which will later dissolve) and some don't: they become part of small associations of stars formed in neighboring 
regions almost by chance. In fact, the observations of  de Wit et al. (2004, 2005), that
are used as standard references for the view that clusters form first, find that 30\% of young massive Galactic 
field stars are not members of clusters or OB associations. Of these, about 50\% are runaway
star \emph{candidates}. They conclude that 4$\pm2\%$\ of the stars in their sample result from
truly isolated high-mass star formation, a number that 
can be reproduced \emph{``assuming that all stars are formed
in clusters that follow a universal cluster distribution
(by $N_*$) with slope $\beta\sim-1.7$\ down to clusters
with a single member.''} If we consider single star clusters
the statement that \emph{all stars are formed in clusters}
becomes a tautology.  

Our results hint at a {\it strong universality} hypothesis for the IMF where not only the power-law part, but the full function may be universal.   Clearly this claim has profound implications for understanding how stars form and therefore its foundations require considerably more observational work than was available for the tests presented in this paper.

\section{Summary}
\label{summary}

This paper combines four separate results within a single
unified view. The unification is actually a result of Equation~\ref{eqMC}
(that is derived formally in the first section of the paper)
which relates the cluster mass function with the probability
distribution of number of stars in clusters, and the (universal)
stellar IMF. These results are
\begin{itemize}
	\item the IMF of a field stellar population and that
	of the clusters out of which the field was built \emph{can} be
	strictly the same. This result contradicts \citet{kroupa2003},
	and was implicit in the models by \citet{larson1982},
	\citet{oey1998, oey2005}, and \citet{dewit2004};

	\item the observations of the lower end of the
	cluster mass function, as given by \citet{lada2003}, agree
	with the random sampling model presented here if: (a) the
	distribution of clusters by the number of stars they contain
	is a scale-free power law, $n^\beta$, with $\beta$\ between -1.8
	and -2.2; (b) the stellar IMF is independent of $n$ and it is
	given by the Salpeter form;

	\item the observed special mass scale for cluster formation claimed
	by Lada and Lada arises from the arbitrary cut-off in the
	number of stars imposed by them;

	\item  the interpretation of the statistics of the most massive star in clusters
	is a valuable tool to study cluster formation processes as the observations,
	taken at face value, violate the \emph{null hypothesis} represented by the
	random sampling model. Although a proper observational study requires a sample
	including many clusters with only a few members, we believe that the observations
	presented by WK2006 are at worst compelling. Nevertheless, it is argued in the
	present work that the discrepancy is due to systems which are rendered gravitationally
	unstable by the presence of one or more very massive stars. 
\end{itemize}

\acknowledgments 
We would like to thank our anonymous referee whose
comments helped us to improve this work.

\appendix

\section{Size-of-sample effect}
\label{ap:sizeofsample}

The size-of-sample effect arises in many areas of astronomy,
and we have studied it in the context of the distribution
of sizes of super-associations in galaxies~ (Selman and Melnick, 2000).
The formalism developed there can be translated
\emph{mutatis mutandi} to the present context as follows:
Let the whole set of star clusters to be analyzed be denoted by ${\cal C} = \{C_i\}_{i=1}^N$,
where $C_i$\ is the i-th cluster with mass  $M_i$\ and maximum
stellar mass $m^{max*}_i$.
We will assume ${\cal C}$\ to be ordered from the most massive to the least massive cluster,
that is,  $i<j \Rightarrow M_i > M_j$.  Let $M_l=\sum_{i=1}^{l} M_i/l$\ be the
average mass of the $l$\ most massive clusters. From ${\cal C}$\ we will draw $N_S$\
sub--samples, ${\cal S}_j$, of equal mass, $M_l$, defined by
\begin{displaymath}
{\cal S}_j = \{C_i\}_{i=j}^{n_j},
\end{displaymath}
where $n_j$\ is defined by the expression,
\begin{displaymath} \sum_{i=j}^{n_j} M_i = M_l.
\end{displaymath}
Thus, the sub-sample ${\cal S}_j$\ contains the cluster $C_j$\ and the next
 $n_j-1$\ less massive clusters, enough to add up to a total mass equal to $M_l$.
We will assign to each sub-sample~$j$\ two numbers:
$\tilde m^{max*}_j$, and $\tilde M^{avg}_j$, defined as
\begin{eqnarray}
\tilde m^{max*}_j = \max_{C_i\in {\cal S}_j} m^{max*}_i,\nonumber\\
\tilde M^{avg}_j = {1\over{n_{j}-j+1}}\sum_{i=j}^{n_j}M_i.\nonumber
\end{eqnarray}
$\tilde m^{max*}_j$\ is equal to the maximum stellar mass of all the members of ${\cal S}_j$,
and $\tilde M^{avg}_j$\ their average mass. We will refer to sub--sample~j
as ``super--cluster''~j.
Because all ``super--clusters'' thus defined
have approximately equal total mass ($\approx M_l$), we can compare
the mass of their maximal star without having a size--of--sample effect.

\begin{figure}
   \epsscale{.80}
   \plotone{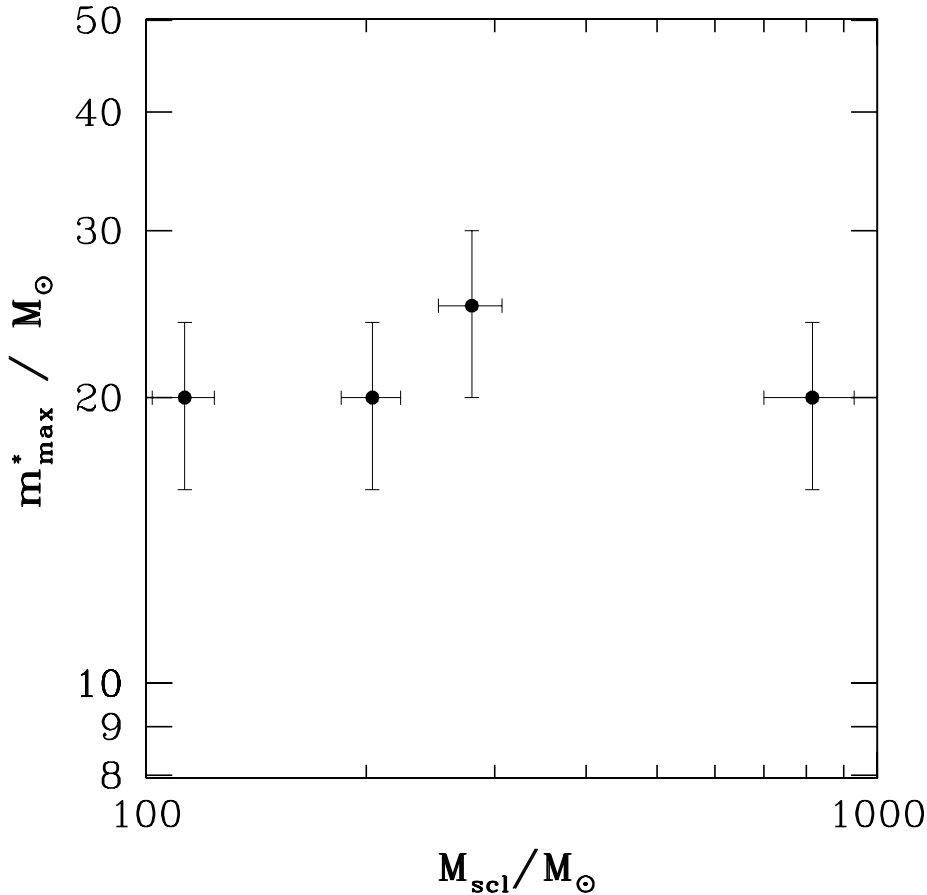}
   \caption{\label{WK2006subsample}
        The maximum stellar mass in a sub-sample of the
        data of WK2006 analyzed as detailed in the main
        text. The x-axis show the average mass of the clusters
	in the 800~M$_\odot$~``super clusters'',
        while the y-axis show the mass of the maximal mass
        star in the super clusters.
           }
\end{figure}

Regrettably, the data set and the metod of analysis used by WK2006 is far from what is needed for this
kind of analysis. Their Table~1 lists 17 clusters with
masses ranging from $25 M_\odot$~to $10^5 M_\odot$. We have seen that we should actualy work with the
number of stars instead of the mass of the clusters as conditioning to cluster mass introduces
unphysical mass scales. If we use an average stellar mass of $0.3M_\odot$\ then the smallest
mass cluster corresponds to a cluster with $\approx$~75 stars while the largest mass
clusters corresponds to $\approx3\times10^5$\ stars.
The sample should include at least 4000 clusters with 75 stars
to meaningfully compare the maximal mass of this artificial ``super-cluster" with the maximal 
mass of a cluster of $10^5 M_\odot$ (such as R136).    Figure~\ref{WK2006subsample}  plots the maximal mass against 
total mass for the best sub-sample of ``super-clusters" that can be constructed from the data of WK2006 using the algorithm described above. This sample consists of NGC6530 as the first ``super-cluster''; 
NGC~2264, Mon R2, and $\sigma$~Ori in the second;   Mon R2, $\sigma$~Ori, NGC~2024, and IC~348 in the third; and $\sigma$~Ori, NGC~2024, IC~348, $\rho$~Oph, NGC~1333,
Ser SVS2, and Taurus-Auriga in the fourth. Each of these four ``super-clusters" has a
total mass of approximately 800~M$_\odot$.  The ``best-set'' shows no correlation between maximal star mass and cluster richness: the claimed correlation was a size-of-sample effect. (It is possible to construct other 
sub-samples having  a more massive star in the upper mass bin, but these sub-samples contain only  2 or  3 super-clusters.)

\bibliography{ms}

\clearpage
\end{document}